\documentclass[tradiabstract]{aa} 
\pdfoutput=1

\usepackage{graphicx}
\usepackage[caption=false,position=top,font=scriptsize,labelformat=empty,aboveskip=0pt,captionskip=0pt]{subfig}
\usepackage{amssymb}
\usepackage{float}
\usepackage{threeparttable}
\usepackage{tikz}
\usetikzlibrary{arrows,shapes,backgrounds}
\usepackage{natbib}
\bibpunct{(}{)}{;}{a}{}{,}
\def\simgreat{\mathbin{\lower 3pt\hbox
     {$\rlap{\raise 5pt\hbox{$\char'076$}}\mathchar"7218$}}}
\def\simless{\mathbin{\lower 3pt\hbox
     {$\rlap{\raise 5pt\hbox{$\char'074$}}\mathchar"7218$}}}

\newcommand{\taueff} {\tau_{\mathrm{eff}}}

\newcommand{\Rve}{R_{\mathrm{V,eff}}}

\defcitealias{2007ApJ...663..320F}{FM07}
\defcitealias{1984ApJ...287..228N}{NP84}
\defcitealias{1998A&A...340..103W}{W98}

\begin{document}

\title{Extinction and dust properties in a clumpy medium}

\titlerunning{Extinction in a clumpy medium}

		\author {P. Scicluna\inst{1,2} \and R.~Siebenmorgen\inst{1}}
\institute{
        European Southern Observatory, Karl-Schwarzschild-Str. 2,
        D-85748 Garching b. M\"unchen, Germany
\and
ITAP, Universit\"at zu Kiel, Leibnizstr. 15, 
24118 Kiel, Germany
}
\offprints{pscicluna@astrophysik.uni-kiel.de}

\date{Received xxx,xxx / Accepted xxx, xxx}

\abstract { 
The dust content of the universe is primarily
  explored via its interaction with stellar photons, which are
  absorbed or scattered by the dust, producing the effect known as
  interstellar extinction.  However, owing to the physical extension of
  the observing beam, real observations may detect a significant
  number of dust-scattered photons.  This may result in a change in
  the observed (or \textit{effective}) extinction with a dependence
  on the spatial distribution of the dust and the spatial resolution
  of the instrument.
We investigate the influence of clumpy dust distributions on the
effective extinction toward both embedded sources and those seen
through the diffuse ISM.
We use a Monte Carlo radiative transfer code to examine the
effective extinction for various geometries.  By varying the number,
optical depth and volume-filling factor of clumps inside the model for
spherical shells and the diffuse ISM, we explore the evolution of the
extinction curve and effective optical depth.
Depending on the number of scattering events in the beam, the
extinction curve is observed to steepen in homogeneous media and
flatten in clumpy media.  As a result, clumpy dust distributions are
able to reproduce extinction curves with arbitrary $\Rve$, the effective ratio of total-to-selective extinction.  The flattening is also able to `wash out' the 2175\,\AA\,bump and results in a shift of the peak to shorter wavelengths.  The mean
$\Rve$ of a shell is shown to correlate with the optical depth of an
individual clump and the wavelength at which a clump becomes optically
thick. 
Similar behaviour is seen for edge-on discs or tori. However, at grazing 
inclinations the combination of extinction and strong forward scattering
results in chaotic behaviour. Caution is therefore advised when attempting
to measure extinction in, for example, AGN tori or toward SNIa or GRB afterglows.
In face-on discs, the shape of the scattered continuum is observed to change
significantly with clumpiness, however, unlike absorption features, individual 
features in the scattering cross-sections are preserved.
Finally, we show that diffuse interstellar extinction is not significantly 
modified by scattering on distance scales of a few kpc.

}

 \keywords{
   Radiative transfer -- dust, extinction -- circumstellar matter -- ISM: structure -- scattering
   }

\maketitle

\section{Introduction\label{sec:intro}}
The presence and nature of dust in the universe can be explored by observing both the thermal radiation it emits, and its influence on stellar photons, which it absorbs and scatters to produce extinction.
Both of these methods sample different dust populations, with emission being most sensitive to the hottest dust components along the entire line of sight, while extinction is sensitive to the full column of dust between the observer and the extinguished source.
Therefore, the wavelength dependence of interstellar extinction can be interpreted in terms of the wavelength dependence of the probability for dust and radiation to interact, i.e. the dust cross-sections. 
{ Extinction is observed to vary on different galactic lines of sight (e.g. \citealt{1990ApJS...72..163F,2007ApJ...663..320F}, hereafter \citetalias{2007ApJ...663..320F}), and extragalactically \citep{1983MNRAS.203..301H,1984A&A...132..389P,1994ApJ...429..582C}.}
As a result, attempts are frequently made to analyse the composition of dust on given lines of sight by fitting the extinction curve using extinction cross-sections for likely mixtures of materials and particle sizes. 
One must, therefore, ensure that all possible biases and systematic effects are accounted for in the treatment of extinction. 

One key and often overlooked bias is the real angular extent of the observing beam in which extinction measurements are made.
Since observations do not use a pencil beam, there is a non-zero probability of detecting scattered light \citep{1972ApJ...176..651M,2009A&A...493..385K}, both increasing the total detected flux and altering the wavelength dependence of extinction.
It is also possible that an inhomogeneous dust distribution will present paths with different optical depths, with the relative covering fractions of the different phases influencing the detected flux.

In galactic observations of the diffuse ISM the impact of scattering
is typically assumed to be negligible, an assumption that we
consider more carefully in Sect. \ref{sec:ism}. Nevertheless, in 
regions where dust and stars are well
mixed or where the physical size of the observing beam is large
compared to the structure of the dusty medium, the fraction of
scattered light can become significant.  This may occur in more
distant galactic star-forming regions \citep{1984ApJ...287..228N} or
for stars embedded in a compact (compared to the resolution) envelope
or disc, i.e. dust-enshrouded (young or evolved) stars
\citep{1996A&A...312..243V,1998A&A...340..103W,2006ApJ...636..362I}.
Similarly, inhomogeneity and scattering effects also become
significant in extragalactic astronomy
\citep{1988ApJ...333..673B,1994ApJ...429..582C,2000ApJ...528..799W},
where an entire star-forming complex can comfortably fit within a
single resolution element.

As a result, unresolved observations of such systems must correctly account for these effects, or they will derive significantly different extinction laws that do not necessarily indicate any change in the physical nature of the dust grains.
Such effects can include both steepening \citep{2009A&A...493..385K} and flattening \citep{1984ApJ...287..228N} of the extinction curve, under- or overestimation of stellar luminosities, or even negative extinction depending on the distribution of the dust and the size of the aperture \citep{2009A&A...493..385K}.

In this paper we make use of numerical radiative transfer models to investigate the effect of scattering and clumpiness on extinction. 
In Sect. \ref{sec:back} we review the relevant theory and previous findings, and Sect. \ref{sec:MC} outlines the computational methods we employ. 
The remainder of the paper then investigates these effects with particular attention paid to circumstellar shells{ , discs} and the diffuse ISM. 

\section{Effective extinction\label{sec:back}}

Following \citet{2009A&A...493..385K} we define the interstellar extinction law 

\begin{equation}
\frac{\tau\left(\lambda\right)}{\tau_\mathrm{V}} = \frac{K_{\mathrm{ext}}\left(\lambda\right)}{K_{\mathrm{ext}}\left(\mathrm{V}\right)},
\label{eqn:ext}
\end{equation}
where $\tau$ is the optical depth and $K_\mathrm{ext}$ the extinction
cross-sections of dust, for observations with infinite resolution.
The so-called true extinction is therefore influenced only
by the column density of extinguishing material (i.e. ISM dust) along
the line of sight and the wavelength dependence of its interactions
with light.  Using the other standard definitions for colour excess
$E\left(B-V\right) = A\left(\mathrm{B}\right) - A_{\mathrm{V}}$ and
$E\left(\lambda -V\right) = A\left(\lambda\right) - A_{\mathrm{V}}$,
where $A$ denotes the extinction in magnitudes, one arrives at

\begin{equation}
k\left(\lambda - V\right) = \frac{E\left(\lambda -V\right)}{E\left(B-V\right)},
\end{equation}
which is the traditional form of the extinction law in terms of colour excess.
This then naturally leads to the definition of the ratio of total-to-selective extinction,
\begin{equation}
R_{\mathrm{V}} = -k\left(0-V\right) =
\frac{\tau_\mathrm{V}}{\tau_\mathrm{V} - \tau\left(\mathrm{B}\right)}.
\label{eqn:RV}
\end{equation}
As a result, the broadband behaviour of the extinction curve can be described to first order by this quantity, $R_{\mathrm{V}}$, and hence so can the dust properties.
Changes in  $R_{\mathrm{V}}$ therefore indicate changes in the dust, usually assumed to result from changes in grain size, as this is, to first order, the dominant factor in the broad-band behaviour of the dust cross-sections.
By finding combinations of dust grains whose cross-sections reproduce the observed constraints, one can therefore hope to understand the composition of dust along a particular line of sight.

To do this, one must assume some dust constituents,
typically some combination of silicon- and carbon-bearing species,
which may be in distinct grain types (e.g. separate carbon- and
silicate-bearing grains) or mixed together (composite grains).  One
must also choose a grain geometry (e.g. spherical, spheroidal,
fractal etc.) and structure (e.g.
homogeneous or porous).  Then, by assuming a size distribution of the
particles, one can compute the extinction cross-sections, albedo,
phase function, etc. for the dust model and compare the wavelength
dependence of these properties to those observed for interstellar
dust.  For a more detailed discussion of the processes involved in
fitting the extinction curve, please refer to the literature
\citep[e.g.][]{2001ApJ...548..296W,2003ARA&A..41..241D,2004ASPRv..12....1V,2012JQSRT.113.2334V,2014A&A...561A..82S}.

However, in real observations a number of effects can complicate the picture.
Firstly, although the extinction cross-sections are defined as 
\begin{equation}
K_{\mathrm{ext}} = K_{\mathrm{abs}} + K_{\mathrm{sca}}
\end{equation}
in general scattering is not isotropic, meaning that observationally
there is a degeneracy \citep{2002ocd..conf....1V} between the albedo $\omega = K_{\mathrm{sca}}
/ K_{\mathrm{ext}}$ and the anisotropy parameter

\begin{equation}
g = \langle cos\left(\theta\right)\rangle =
  \int p\left(\cos\left(\theta\right)\right)
  \cos\left(\theta\right)\mathrm{d}\cos\left(\theta\right)\label{eqn:gfac}
\end{equation}

\noindent where $p\left(\cos\left(\theta\right)\right)$ is the probability density function of the cosine of the scattering angle $\theta$, which parametrises the expectation of the scattering
direction, with 1 corresponding to pure forward scattering
and -1 to pure back-scattering.

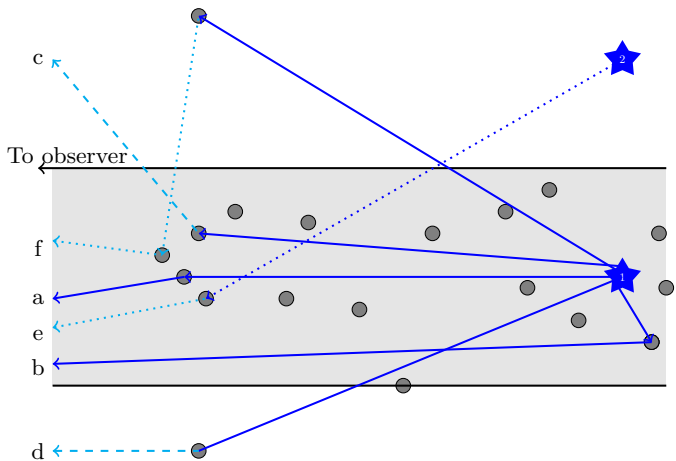
\begin{figure}
\resizebox{\hsize}{!}{
		\begin{tikzpicture}
			\fill [gray,opacity=0.2] (0.2,4.5) rectangle (8.6,7.5);
			\draw [fill=gray] (2,6) circle (0.1);
			\draw [fill=gray] (2.2,6.6) circle (0.1);
			\draw [fill=gray] (2.3,5.7) circle (0.1);
			\draw [fill=gray] (1.7,6.3) circle (0.1);
			\draw [fill=gray] (2.2,9.6) circle (0.1);
			\draw [fill=gray] (8.5,6.6) circle (0.1);
			\draw [fill=gray] (8.6,5.85) circle (0.1);
			\draw [fill=gray] (8.4,5.1) circle (0.1);
			\draw [fill=gray] (6.4,6.9) circle (0.1);
			\draw [fill=gray] (7.4,5.4) circle (0.1);
			\draw [fill=gray] (3.4,5.7) circle (0.1);
			\draw [fill=gray] (4.4,5.55) circle (0.1);
			\draw [fill=gray] (5.4,6.6) circle (0.1);
			\draw [fill=gray] (6.7,5.85) circle (0.1);
			\draw [fill=gray] (7.,7.2) circle (0.1);
			\draw [fill=gray] (3.7,6.75) circle (0.1);
			\draw [fill=gray] (5.,4.5) circle (0.1);
			\draw [fill=gray] (2.7,6.9) circle (0.1);
			\draw [fill=gray] (8.4,5.1) circle (0.1);
			\draw [fill=gray] (8.4,5.1) circle (0.1);
			\draw [fill=gray] (2.2,3.6) circle (0.1);
			\node[star,star point ratio=1.8, fill=blue,minimum width=1mm,scale=0.5,text=white] at (8,6) {1};
			\draw [color=black,thick] (0.2,4.5) -- (8.6,4.5);
			\draw [color=black,<-,thick] (0,7.5) -- (8.6,7.5);
			\node at (0.4,7.68) {To observer};
			\draw [color=blue,<-,thick] (0.2,5.7) -- (2,6);
			\node at (0,5.7) {a};
			\draw [color=blue,<-,thick] (2,6) -- (7.95,6);
			\draw [color=blue,<-,thick] (2.2,6.6) -- (7.95,6.15);
			\draw [color=blue,<-,thick] (2.2,9.6) -- (7.95,6.075);
			\draw [color=blue,<-,thick] (8.4,5.1) -- (7.95,5.85);
			\draw [color=blue,<-,thick] (0.2,4.8) -- (8.4,5.1);
			\node at (0,4.75) {b};
			\draw[color=cyan,<-,thick,dotted](1.7,6.3) -- (2.2,9.6);
			\draw[color=cyan,<-,thick,dotted] (0.2,6.5) -- (1.7,6.3);
			\node at (0,6.4) {f};
			\draw [color=cyan,<-,thick,dashed] (0.2,9) -- (2.2,6.6);
			\node at (0,9) {c};
			\draw[color=blue,<-,thick](8,6) -- (2.2,3.6);
			\draw[color=cyan,<-,thick,dashed]  (0.2,3.6) -- (2.2,3.6);
			\node at (0,3.6) {d};
			\node[star,star point ratio=1.8, fill=blue,minimum width=1mm,scale=0.5,text=white] at (8,9) {2};
			\draw[color=blue,<-,thick,dotted] (2.3,5.7) -- (8,9);
			\draw[color=cyan,<-,thick,dotted] (0.2,5.3) -- (2.3,5.7);
			\node at (0,5.2) {e};
		\end{tikzpicture}}
		
		\caption{Scattered photons may still be observed on the detector. 
		The grey-shaded region between the two black lines indicates the volume swept out by the observing beam of the telescope as it extends into space.
		Photons (light and dark blue lines) that arrive at the detector at the end of this region (marked `To observer') will be observed as though they originated at the star.
		The grey circles represent a distribution of dust along the line of sight toward the star being observed.
		The dark full lines show the contributions we consider here: `undeflected' photons (a) which are forward scattered and do not leave the beam, and photons which are back scattered (b) into the beam. 
		The pale dashed lines (c,d) show cases where the scattering event leads to the photon leaving the observing beam.
		Finally, the dotted lines (e,f) represent cases that may contribute to observations, but occur with significantly lower probabilities and are hence not considered in this paper. 
		}

		\label{fig:scatter}
		
\end{figure}

Furthermore, in real astronomical observations, the aperture or beam
in which the extinction is measured is not a pencil beam and has some
physical extension, determined by the resolution.  Hence, unresolved
structure within the beam can alter the observed extinction by for
example
\begin{itemize}
\item partially occulting the source;
\item inhomogeneities biasing observations toward low-$\tau$ paths;
\item scattering light into the beam;
\end{itemize}
which can combine with the aforementioned degeneracy.
Figure \ref{fig:scatter} depicts this in cartoon fashion { for sources in the far field}.

When the extinguished source and the observer are roughly equidistant
from the extinguishing material, this effect is negligible
\citep{2009A&A...493..385K}, but is increasingly significant the
shorter the physical distance between the star and the attenuating
matter. It naturally follows that this effect is most significant for
embedded objects and extragalactic observations.

To account for this, previous authors \citep[see e.g.][]{2009A&A...493..385K} have defined the \textit{effective} optical depth and extinction curve i.e.
\begin{equation}
	\frac{\taueff\left(\lambda\right)}{\tau_\mathrm{V,eff}} \neq \frac{K_{\mathrm{ext}}\left(\lambda\right)}{K_{\mathrm{ext}}\left(\mathrm{V}\right)},
\label{eqn:effext}
\end{equation}
where $\taueff$ is the optical depth one derives from the observations; i.e. the negative of the logarithm of the ratio of the observed flux to the flux that \emph{would be observed in the absence of dust} \begin{equation}
\taueff = - \mathrm{ln}\frac{F_{\mathrm{obs}}}{F_{\mathrm{0}}} .
\label{eqn:taueff}
\end{equation}
 From this follows the definitions of $\Rve$ as in equations
 \ref{eqn:ext} to \ref{eqn:RV} with $\taueff$ instead of $\tau$.  This
 is similar to the definition of attenuation optical depth $\tau_\mathrm{att}$ used in
 e.g. \citet{2000ApJ...528..799W}, but noticeably different from the
 definitions of $\taueff$ used in \citet{1996ApJ...463..681W} and
 \citet{1998A&A...340..103W} and $\tau_\mathrm{att}$ in \citet{2005ApJ...619..340F}, which exclude the contribution from scattered photons.

It should also be clear that unlike $\tau$, $\taueff$ is a function not only of the source and its dust distribution, but also of the aperture in which it is observed \citep{2009A&A...493..385K}.
\citet{2009A&A...493..385K} also emphasises that $\taueff$ is never larger than $\tau$, and that it can even be negative (e.g. in a reflection nebula).
When the dust distribution is homogeneous, $\taueff$ depends only on $\tau$, the dust composition and the aperture, while for inhomogeneous media the spatial distribution of dust and the viewing angle are clearly also important \citep{1998A&A...340..103W}.

\section{Monte Carlo models\label{sec:MC}}

As the exploration of the effective extinction necessitates accurate radiative transfer modelling in inhomogeneous media, we must use Monte Carlo methods.

We use an implementation originally described in
\citet{2008ipid.book.....K}, and significantly expanded upon in
\citet{2012A&A...539A..20S} and \citet{2012ApJ...751...27H}. Our code
allows for an arbitrary choice of geometry, dust composition, and
illumination source, and includes anisotropic scattering.  By
launching packets of radiation from the source and following their
interactions with the surrounding dust distribution we solve the
radiative transfer.

\begin{figure}
\resizebox{\hsize}{!}{\includegraphics[scale=0.5,clip=true,trim=0.3cm 0.5cm 4.5cm 16.5cm]{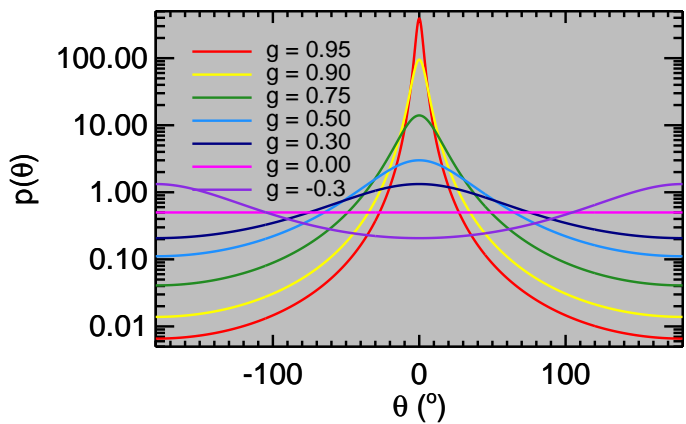}}

\caption{Probability density function of the Henyey-Greenstein phase function as a function of scattering angle for a representative range of g-factors. $\theta = 0\degr$ indicates that the outgoing direction of the scattered photon is identical to that of the incoming one, while $\theta = \pm 180\degr$ implies a reversal of direction { (back scattering).}}
\label{fig:HGfun}
\end{figure}

The dust distribution consists of a Cartesian grid of densities and
temperatures.  The interactions of the radiation packets are then
computed based on the method in \citet{2008ipid.book.....K} which
employs the `immediate temperature update' method of
\citet{2001ApJ...554..615B}.  We have extended this method to include
the \citet{1999A&A...344..282L} algorithm for the dust temperatures in
optically thin regions to reduce the uncertainty in the dust
temperatures, and to include anisotropic scattering by sampling
scattering angles from the Henyey-Greenstein (HG) phase function
\citep{1941ApJ....93...70H}
\begin{equation}
p\left(\cos\left(\theta\right)\right) = \frac{1}{4\pi} \frac{1 - g^2}{\left(1 + g^2 - 2 g \cos\left(\theta\right)\right)^{3/2} },
\label{eqn:HG}
\end{equation}
where $g$ is the anisotropy parameter (Eq. \ref{eqn:gfac}) derived
from Mie calculus \citep{Mie1908,1983asls.book.....B}.  This can be
re-arranged to give
\begin{equation}
\cos\left(\theta\right) = \frac{1}{2g} \left[1 + g^2 - \left(\frac{1 - g^2}{1+g\left[2P-1\right]}\right)\right]
\label{eqn:invHG}
\end{equation}
where $P=\int p\left(\cos\left(\theta\right)\right)
\mathrm{d}\cos\left(\theta\right)$ is the cumulative probability
distribution, from which the scattering angle can be sampled directly.
As this function contains a singularity for $g=0$ it is necessary to
treat these cases separately by explicitly interpreting them as isotropic
scattering.  Figure \ref{fig:HGfun} shows probability density
functions $p\left(\theta\right)$ for the HG function over a
representative range of the $g$ parameter.

\begin{figure}
\resizebox{\hsize}{!}{\includegraphics[scale=0.5,clip=true,trim=0.25cm 1.8cm 1.25cm 9.25cm]{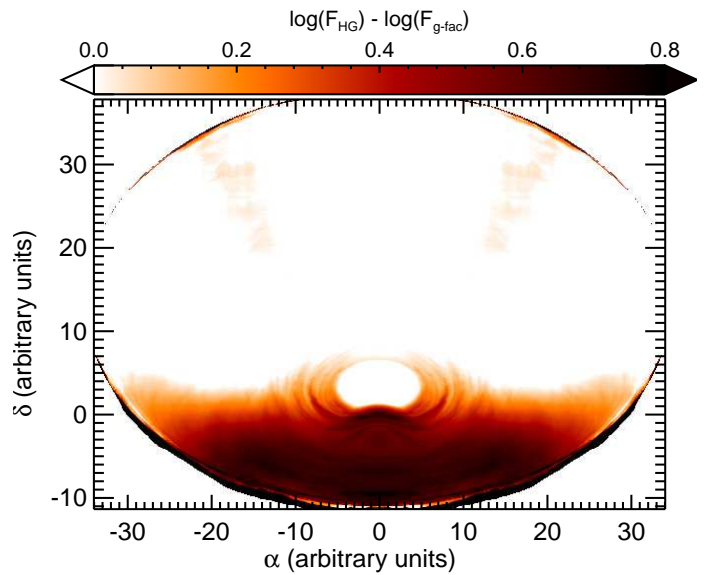}}
\caption{Difference image between the V-band scattered flux computed assuming the HG phase-function ($F_\mathrm{HG}$) and a pseudo-isotropic one ($F_\mathrm{g-fac}$), computed for a dust disc with a half-opening angle of $30\degr$ viewed from an angle of $45\degr$ from the rotation axis using {  a dust model composed of amorphous carbon and silicates}. The region of the disc at $\delta \leq 0$ is the near-side of the disc. The HG function reproduces the strength of forward scattering much more effectively, with differences between the two methods of up to an order of magnitude, although the integrated scattered flux is the same using both methods. }
\label{fig:scadif}
\end{figure}

The importance of anisotropic scattering {  is demonstrated} in
Fig. \ref{fig:scadif}, which shows a difference image comparing the
same model viewed in scattered light using either the HG function or a
pseudo-isotropic approximation, in which the scattering cross-sections
are reduced by $K^{\prime}_{sca} = \left(1-g\right)K_{sca}$, which
effectively divides the scattering into an isotropically-scattered
component and a forward-scattered (unscattered) component, which works
well for $0\leq g \ll 1$ and $g=1$ but becomes increasingly poor as
$\left| g \right| \rightarrow 1$.  It is clear that the HG function
shows a completely different distribution of scattered flux, with the
near-side of the disc significantly brighter and the far-side
darkened.

\begin{figure}
\resizebox{\hsize}{!}{\includegraphics[scale=0.5,clip=true,trim=1.5cm .5cm 3.75cm 16.5cm]{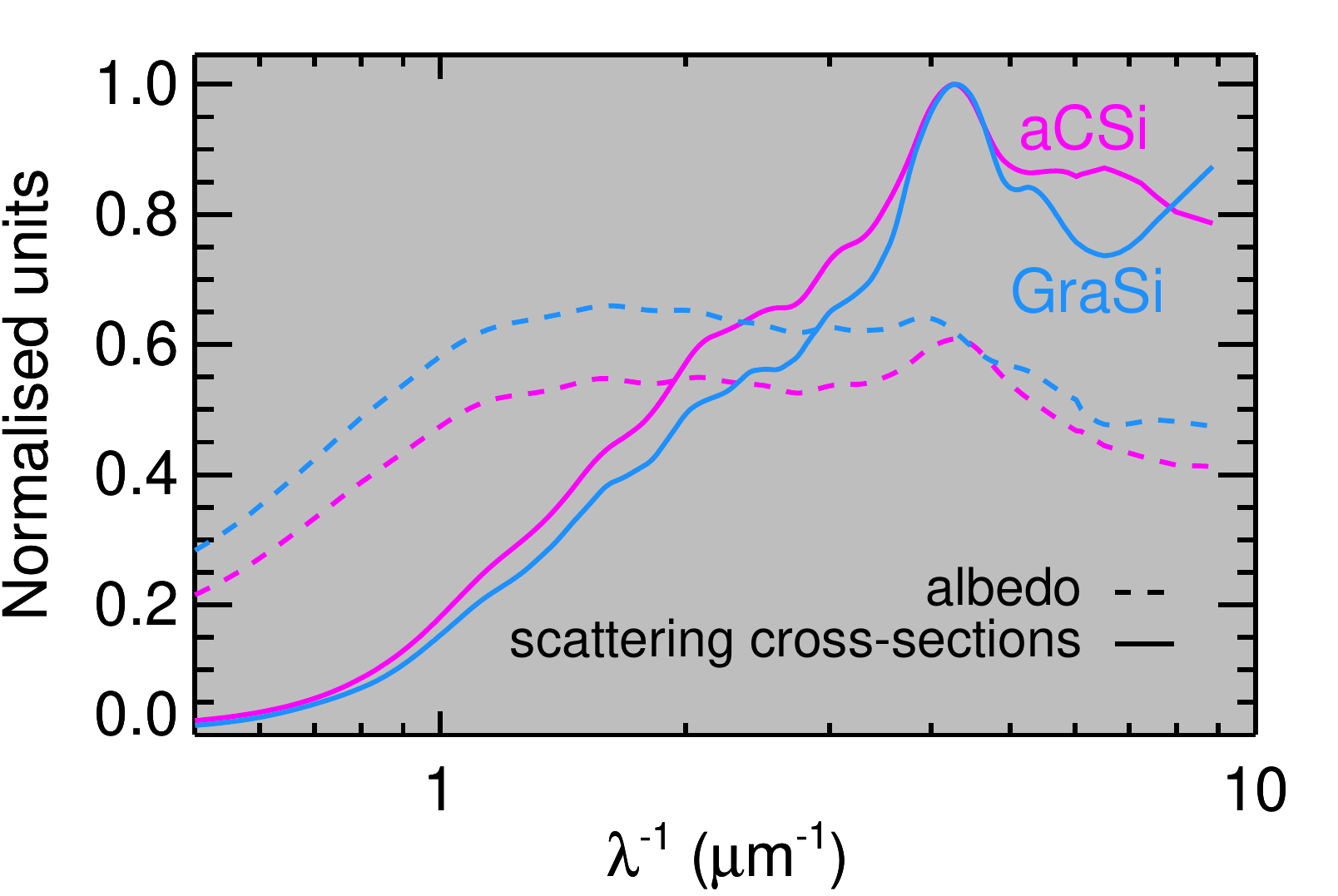}}

\caption{Normalised scattering cross-sections (full lines) and albedo (dashed lines) as a function of wavelength for both dust models. While the albedo is rather flat, the cross-sections show a strong peak between $200-250\mbox{ nm}$.}
\label{fig:ksca}
\end{figure}

Since we are able to follow the packets explicitly we can directly
compute $\taueff$ for all wavelengths, and hence $\Rve $, simply by
counting how many packets emerge from the cloud at a given wavelength.

A number of apertures can be defined on the basis of viewing angle or
physical location.  As photon packets exit the model grid, they are
added to the statistics for the relevant apertures.  We thus build-up
effective extinction curves by computing the number of photons within
the aperture and comparing to the number that would have been detected
in the absence of dust (Eq. \ref{eqn:taueff}).

Our code, including anisotropic scattering, is parallelised using the
OpenMP\footnote{http://openmp.org/} API for use on shared memory machines. When we are only
interested in the influence of scattering and extinction by dust on
the ultraviolet, optical and near infrared we further optimise the
code by neglecting dust emission. In this case all photon packets
absorbed by the dust are discarded and the runtime of the code is
decreased by a factor of four.  Nevertheless the models remain
computationally intensive, and although the physical scale of the
apertures in which the extinction is computed are correct, it is
necessary to overestimate their angular extent {  as seen from the central star} to develop sufficient
statistics without the runtime becoming infeasible.
{  This is done by placing the apertures closer to the star than the assumed distance to the observer.}

In these studies we consider the so-called MRN grain size distribution
($dn\left(a\right) \propto a^{-q} \ da$, $ q = 3.5$
\citet{1977ApJ...217..425M}) of silicate and amorphous carbon.  Since
we are looking for changes caused by the dust geometry, the precise
dust model chosen is not important. We make use of two dust models
depending on the conditions we wish to explore:
	\begin{itemize}
		\item amorphous carbon and silicates (aCSi), using optical constants from \citet{1996MNRAS.282.1321Z} and \citet{2003ApJ...598.1017D}, respectively;
		\item graphite and silicates (GraSi), using optical constants from \citet{2003ApJ...598.1017D}.
	\end{itemize}

Both models consist of carbonaceous grains with radii between 16\,nm
and 130\,nm and silicate grains between 32\,nm and 260\,nm.  The bulk
density of the dust grains is 2.5$\mbox{ g cm}^{-3}$, and the
carbon-to-silicate abundance ratio is 6.5.  The scattering
cross-sections of both models peak around 250\,nm (see Fig.~\ref{fig:ksca}); { this results from highly efficient scattering from grains with size $a \approx\lambda / 2\pi$, i.e. the size of the smallest silicate grains.}

We are also able to generate high signal-to-noise images by
post-processing the output of the radiative transfer simulations by
using a ray-tracer.  Scattered light images require that we first
store the position, frequency and direction of photons before a
scattering event. Then this information is read into the ray-tracing
algorithm, and used to calculate the angle between the incident photon
and the observing direction
\begin{equation}
\cos\theta = \vec{\hat{e}_i} \cdot \vec{\hat{e}_o}
\end{equation}
 where $\vec{\hat{e}_{i,o}}$ indicate the incoming and observer
 direction unit vectors, respectively.  We then determine the
 probability of scattering the photon packet into the viewing direction
 from the scattering phase-function (Eq. \ref{eqn:HG}), and this
 fraction of the packet is added to the ray.  This is similar to the
 so called peel-off technique \citep{1984ApJ...278..186Y}.
 Emission images are computed by integrating the emission determined
 from the dust temperatures, cross-sections and optical depth along
 the line of sight.  Both routines include a correction for the
 attenuation caused by the line-of-sight optical depth. In the case of
 a very small aperture (e.g. simulated observations of extinction in
 the diffuse ISM) the same signal-to-noise ratio can be achieved in a
 much shorter time by exploiting this capability { to integrate over all scattering events.}
 
 { 
 While astronomical ray-tracing applications usually only consider models in the far field, allowing them to use parallel rays, models of the ISM need to consider photon scattering along the entire line of sight.
 Hence, we apply perspective-projection ray tracing \citep[e.g.][]{Appel} to capture the effect of the beam widening as the distance from the observer increase; the direction of a ray now depends upon its position on the detector, and all rays are divergent.
 This requires us to update the prescription in \citet{2012ApJ...751...27H}.
 
 In order to determine the deflection of each ray, we must first know the field of view required from the image.
 This is calculated from the inverse tangent of the projected size of model and the distance to the object e.g. for a cuboid where the long ($z$) axis is parallel to the central ray
 \begin{equation}
 \theta_{FOV} = 2 \tan^{-1}\left(\frac{\Delta x}{D}\right)
 \end{equation}
 where $\Delta x$ is the length of the $x$-axis of the model and $D$ is the distance from the observer to the object.
 This ensures that the entire model fits into the image at the location of the object.
 
 The deflection between adjacent pixels is then given by $\delta\theta_{pix} = \theta_{FOV}/n_{\rm pix}$ for a square image with $n_{\rm pix} \times n_{\rm pix}$ pixels.
 The direction of the ray launched from each pixel can then be found by rotating the direction of the vector joining the detector to the object by integer multiples of $\delta\theta$.
 
 Because the rays are divergent, the size of each pixel becomes a function of the distance from the detector along the ray.
 If $d$ is the distance the ray has travelled so far, then the pixel area $A\left(d\right) = \left(d \delta\theta\right)^2$.
 This value must be substituted for the constant value of the pixel size $A$ in Eq. 17 of \citet{2012ApJ...751...27H}.
 The algorithm is otherwise identical to standard parallel-projection ray-tracing methods.
 }  

We use the Monte Carlo code to calculate the temperature structure and
distribution of scattering events in the model cloud and then
calculate images at all the wavelengths of interest ($\lambda < 3\mu
m$) with the ray--tracer.  In an analogous manner to real
observations, these images are then compared to identical images of
dust-free simulations, and the effective optical depth and extinction
curve are calculated (Eq.~\ref{eqn:taueff}).  

\section{Results\label{sec:res}}
	\subsection{Influence of clumps on extinction in circumstellar shells\label{sec:shell}}

We wish to study the influence of clumps on the effective extinction curve, and so first benchmark the results of our treatment by comparing them to examples as found in the literature.
Therefore, we compute the effective extinction curves for clumpy spherical shells, similar to those treated by \citet{1998A&A...340..103W}.  
In our case each clump occupies one cell of the model grid. 
This grid consists of a cube containing $\left[ nx, ny, nz\right] = \left[ 60, 60, 60\right]$ cells of equal size.
The shell is completely described by its inner and outer radii $R_\mathrm{in}$ and $R_\mathrm{out}$, the number of clumps $N_\mathrm{cl}$ and the total dust mass in the shell $M_\mathrm{d}$.  
The range of parameters used is included in Table \ref{tab:shellpar}.
		
		\begin{table}
		\caption{Clumpy shell model parameters}
		\label{tab:shellpar}
		\centering
		\begin{normalsize}
		\begin{threeparttable}
		\begin{tabular*}{\hsize}{l l l l}
		\hline\hline
		\multicolumn{3}{c}{Parameter}  &   Values   \\ \hline
		& & &\\
		Inner radius &[AU] & $R_\mathrm{in}$  & 12 \\
		Outer radius &[AU] & $R_\mathrm{out}$ & 120 \\
		Dust mass\tnote{a} &[$10^{-7} \ M_{\odot}$] & $M_\mathrm{d}$ &
1.8, 5.5 \\
		Optical depth &(aCSi) & $\tau_\mathrm{V}$\tnote{b}\, & 1.0, 3.1 \\
		 &(GraSi) &  & 0.9, 2.9\\
		Number of clumps&& $N_\mathrm{cl}$ & 0\tnote{c}, 500, 1000,  \\
&		 & & 2000, 3000,  \\
&		 & & 5000, 10000 \\
	
		\hline
		\end{tabular*}
		\begin{tablenotes}
		\item [a] Total mass in dust in the shell.
		\item [b] Radial optical depths of the homogeneous shells of the respective dust masses.
		\item [c] 0 corresponds to a homogeneous shell.
		\end{tablenotes}
		\end{threeparttable}
				\end{normalsize}
		\end{table}

		The $N_\mathrm{cl}$ clumps are distributed randomly
                throughout the volume of the shell by selecting cubes
                from the model grid; selected cubes will contain dust,
                and non-selected cubes remain empty\footnote{ To avoid
                  the possibility of infinite loops for high
                  $N_\mathrm{cl}$, if the same cell is selected a
                  second time, it will have its density doubled; if it
                  is selected again it will then have triple the
                  density, \textit{et cetera}.}.
                  The total mass of the shell is then normalised to the input value.
		As a result, we have a distribution of identical clumps of a given total mass.

\begin{figure}
			\resizebox{\hsize}{!}{\includegraphics[clip=true,trim=25cm 8cm 12cm 8cm]{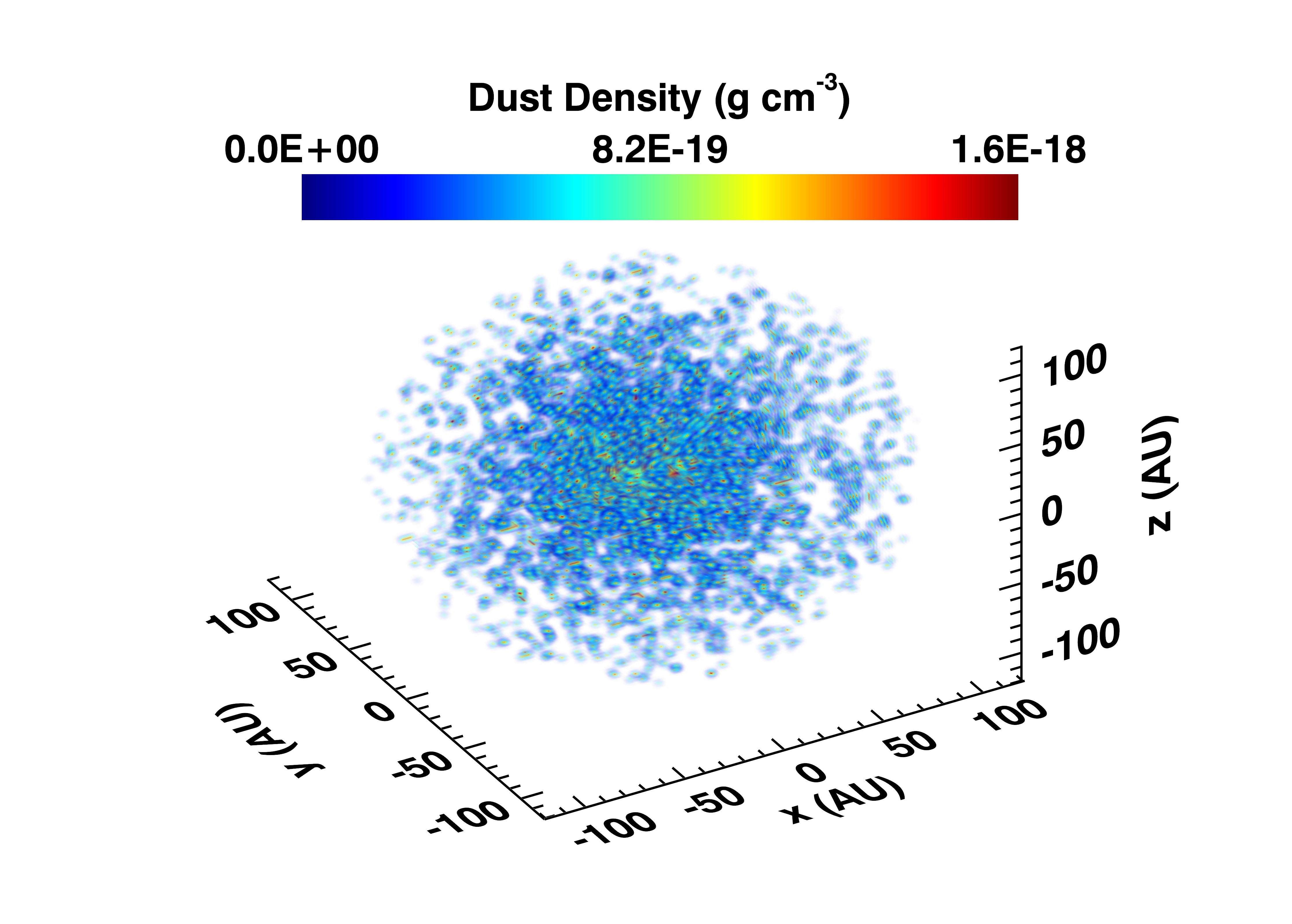}}
			
			\caption{An example of the dust distribution in a clumpy shell shown in the 3D model volume. The source is located at the origin. While the colour indicates the dust density (i.e. number density of grains $\times$ mass of dust grains) at a point, the opacity of the colours is related to the total column density. Upon close inspection it is clear that neighbouring clumps may connect to form filamentary structures. }
			\label{fig:shellclumps}
		\end{figure}

\begin{figure}
		\resizebox{\hsize}{!}{\includegraphics[scale=0.5,clip=true,trim=2.1cm 0.5cm 4.3cm 5cm]{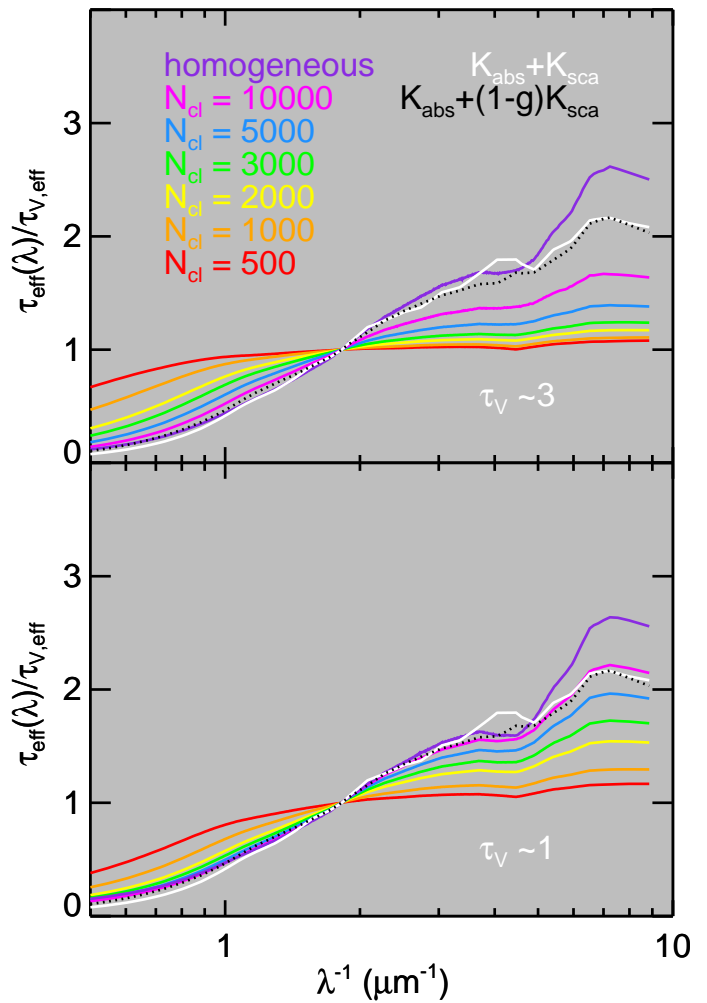}}
		\caption{Effective extinction curves for clumpy circumstellar shells { as a function of N$_{\rm cl}$} using the aCSi model. The line colours correspond to the models indicated in the top left. As the number of clumps decreases, the clumps become more optically thick and the effective extinction curve flattens. {  The white solid line shows the input dust cross--sections normalised to the V band, and the black dotted line the same after a reduction in the scattering efficiencies by $\left(1-g\right)$.} { The two panels refer to different dust masses and hence homogeneous--shell optical depths as indicated on the panel.}}
		\label{fig:SaCcurves}
\end{figure}

\begin{figure}
		\resizebox{\hsize}{!}{\includegraphics[scale=0.5,clip=true,trim=2.1cm 0.5cm 4.3cm 5cm]{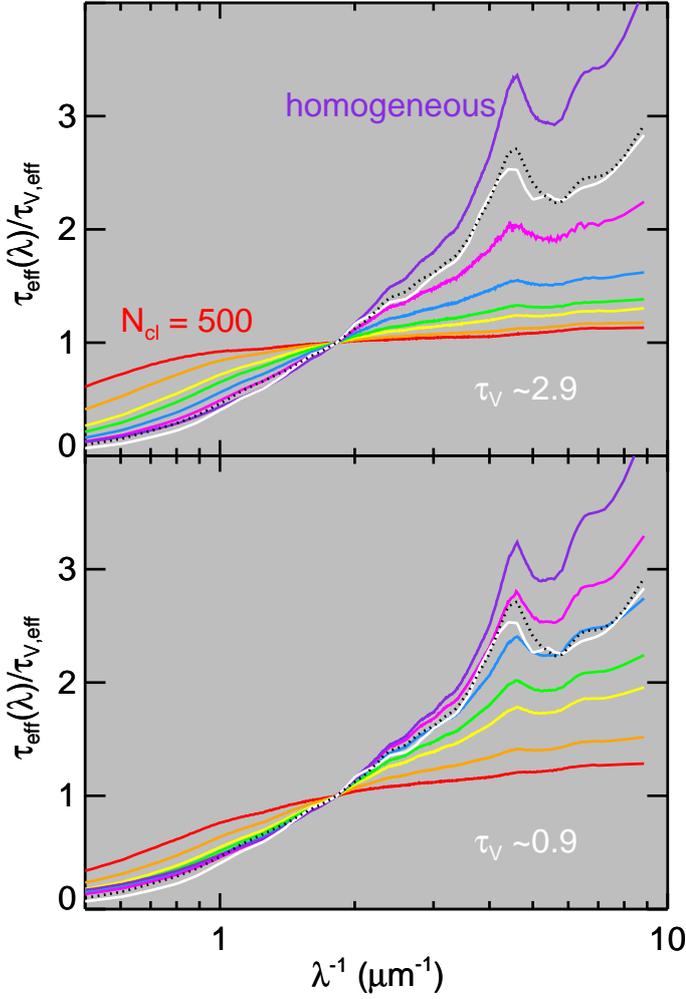}}
		\caption{As in Fig. \ref{fig:SaCcurves} using the GraSi model. The same effects occur with both dust models. In addition it is clear that the 2175\AA\,feature is suppressed as the clump optical depth increases.}
		\label{fig:GraScurves}
\end{figure}

As the clumps are randomly distributed, we must explore a large range
of random seeds\footnote{The random seed is the state used to
  initialise the random number generator's output. By changing this
  value between different models by significantly more than the number
  of random numbers required we obtain nearly independent streams of
  pseudo-random numbers.} for each model to be able to extract average
behaviour, and to quantify the variations that could be seen between
otherwise identical shells.  As changing the distribution of clumps
and changing the angle from which a clumpy shell are equivalent, the
variations between models with different seeds can also be interpreted
in terms of a change in the location of the observer relative to a
fixed axis.  An example of the density distribution of these shells
can be seen in Fig. \ref{fig:shellclumps}.  For comparison, we compute
homogeneous shells where $R_\mathrm{in}$, $R_\mathrm{out}$, and
$M_\mathrm{d}$ are the same as in the clumpy cases.
 As the dust
mass is fixed, models with fewer clumps have clumps of higher optical
depths, which lie in the range $0.1 \leq \tau_\mathrm{cl}
\leq 30${ , where $\tau_\mathrm{cl}$ is the optical depth at V band between two opposite faces of a clump}. The extinction curves are computed by treating each face of the model cube as a large aperture (see Sect.~\ref{sec:MC}), and are shown in
    Figs.~\ref{fig:SaCcurves}--\ref{fig:GraScurves}.
		
		\begin{figure}
		\resizebox{\hsize}{!}{\includegraphics[scale=0.5,clip=true,trim=1cm 0.5cm 4.3cm 5cm]{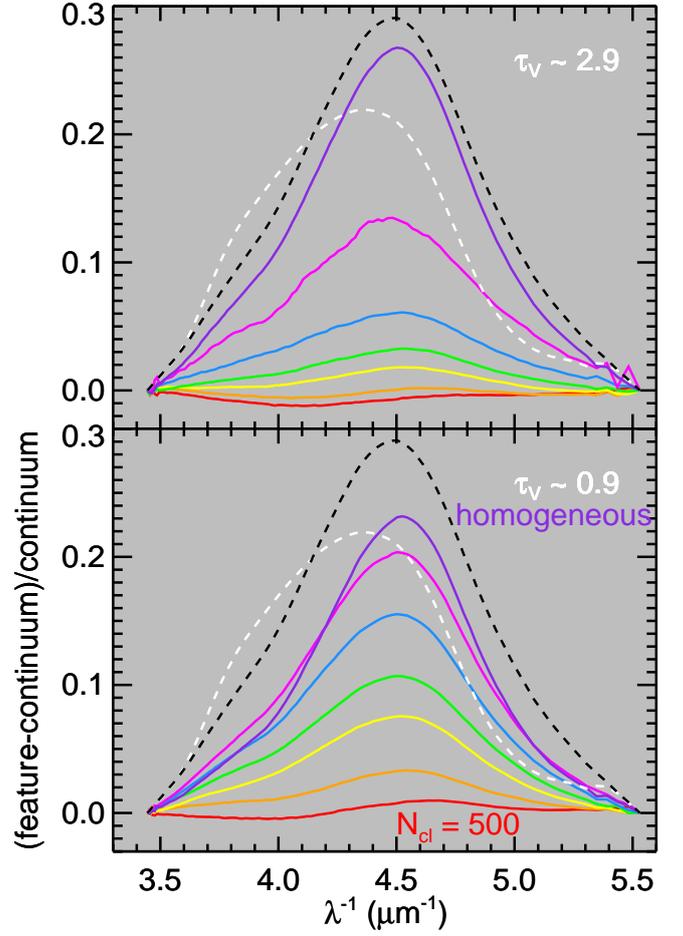}}
		\caption{Fractional strength of the $2175\mbox{\AA}$ feature of the effective exctinction curves shown in Fig. \ref{fig:GraScurves} (colours as in Fig. \ref{fig:SaCcurves}). On top of the suppression of the feature, it is apparent that the shape of the feature is different from that given by the input dust cross-sections. }
		\label{fig:feature}
\end{figure}				
		
		As in \citet{1998A&A...340..103W}, we find that shells that consist of optically thick clumps have generally flatter extinction curves than that given by the dust cross-sections, and in the most extreme cases the extinction curve can become completely grey (Figs.~\ref{fig:SaCcurves}--\ref{fig:GraScurves}), in accordance with \citet{1984ApJ...287..228N}.
		Furthermore, the homogeneous shells (and those with optically thin clumps) have extinction curves that are significantly steeper than that one would derive from the cross-sections, similar to the findings of \citet{2009A&A...493..385K}.
		When using the GraSi dust model to include the 2175\,$\mbox{\AA}$ extinction bump, we see, as \citet{1984ApJ...287..228N}, that as
$\tau_\mathrm{cl}$ increases, not only does the extinction curve
flatten, but as in \citet{1984ApJ...287..228N}, the feature
is weakened and eventually flattened out (Fig.\ref{fig:feature}).

		However, we also notice that in no case does the \emph{shape} of the feature agree with the input dust cross-sections, regardless of whether the cross-sections are parametrised in terms of $K_\mathrm{ext} = K_\mathrm{abs} + K_\mathrm{sca}$ or $K_\mathrm{ext} = K_\mathrm{abs} + \left(1-g\right)K_\mathrm{sca}$.
		{ In particular, the wavelength of the peak of the feature shifts, generally to shorter wavelengths, although there is no clear trend with N$_{\rm cl}$.}
		
		Finally, the wavelength dependence of the extinction at $\lambda\ge 1\mu$m tends toward parallel power-laws i.e. with the same gradient but offset in $\taueff / \tau_\mathrm{V,eff}$ \citep{1984ApJ...287..228N}.
		This may indicate that other indicators of extinction are preferable to those given in the V-band, e.g. normalised to the JHK or even L bands, provided that one is confident that the dust is sufficiently cold to neglect dust  emission in these bands.
		Alternatively, one may be able to use the wavelength at which the infrared extinction deviates from a power-law to infer the optical depth of clumps in the medium. 
		The wavelength at which this deviation occurs appears to be related to the optical depth of the clumps, with more optically thick clumps showing power-law behaviour at longer wavelengths where they become optically thin. 
		
		A number of differences exist between our models and those in the literature.
		\citet{1998A&A...340..103W} integrated the emergent flux over $4\pi$ steradians while we bin the extinction curve into directional apertures; as clumpiness naturally introduces some directionality to the shell, averaging over all directions neglects this.
		Contrary to the models of \citet{1984ApJ...287..228N}, which treated the extinguishing medium as a clumpy screen, the use of a shell geometry results in the inclusion of back-scattering, which requires that directionality be included. 
		\citet{2009A&A...493..385K} on the other hand tailored their models to low optical depth clumps, neglecting the high clump optical depth cases we include here. 
		
	\subsection{Light scattering by clumpy circumstellar discs\label{sec:disc}} 
		
		Having ensured that we reproduce the literature results concerning extinction in clumpy media which arise due to scattering, it may be of interest to consider the influence of clumpiness on the observation of scattered light itself.
		
\begin{figure}
			\resizebox{\hsize}{!}{\includegraphics[clip=true,trim=25cm 8cm 12cm 8cm]{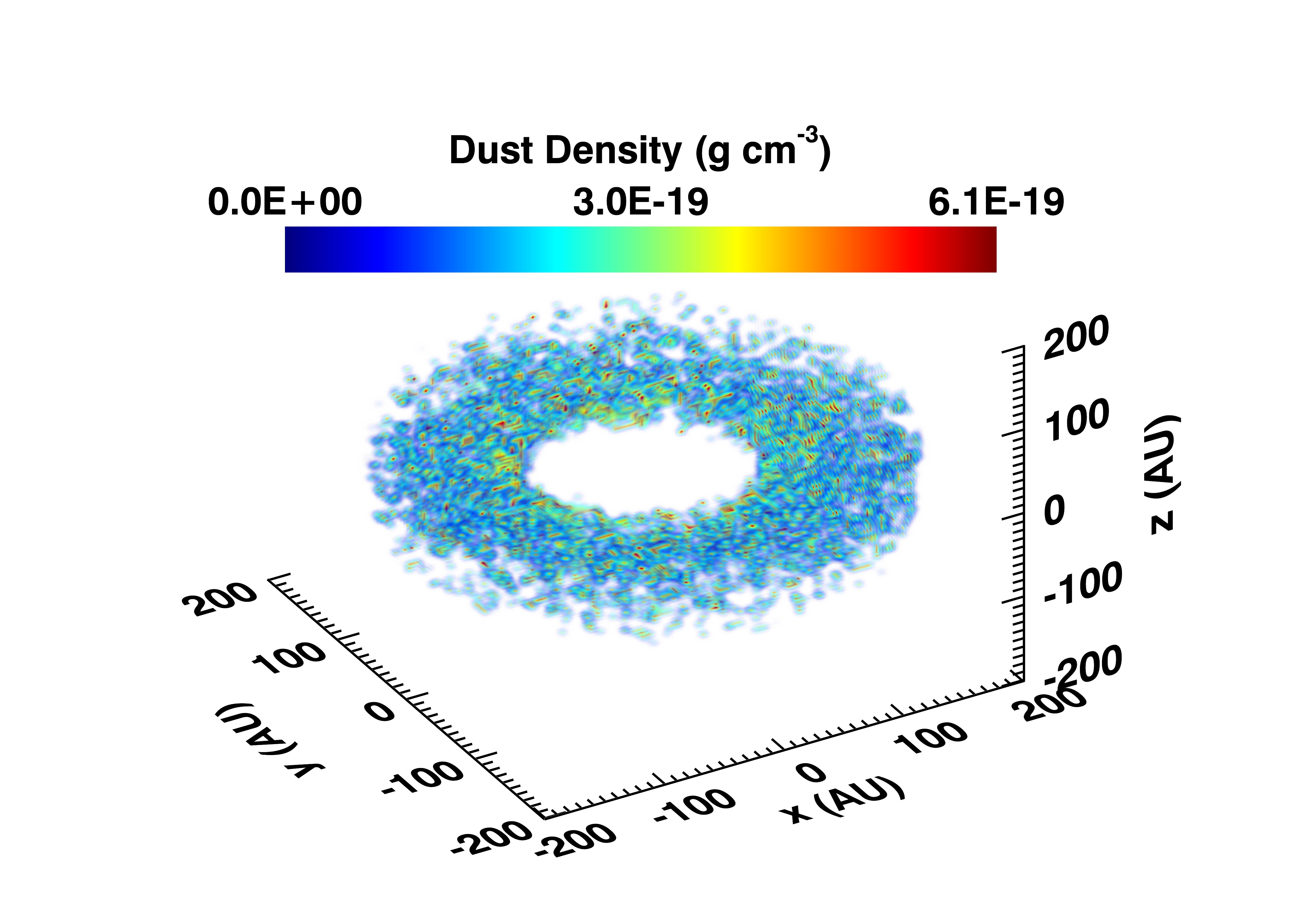}}
			
			\caption{Similar to Fig. \ref{fig:shellclumps} but showing an example of the dust density distribution in a clumpy disc model.}
			\label{fig:discclumps}
		\end{figure}

		To investigate this, we require models in which the view of the source is unobstructed, so that the stellar contribution can be easily subtracted to leave only the scattered photons. 
		We thus model circumstellar discs constructed in a similar manner to the circumstellar shells in Sect. \ref{sec:shell}, but allowing dust only within a given opening angle of the equator; regions above this are dust free.
		We choose their inner and outer radii to approximately match those observed for the Vega outer debris disc (80 and 200 AU, respectively) and include 0.1$M${\scriptsize$_{\bigoplus}$} of dust using the aCSi model. 
		The number of clumps and the opening angle of the disc, { $\alpha$,}\footnote{The disc consists of a dust filled equatorial region whose surface is at height $h=r\tan\alpha$ above the mid-plane, with conical dust-free regions at each pole.} are treated as free parameters. 
		One such example can be seen in Fig. \ref{fig:discclumps}. 
		The effective extinction is then measured for an observer seeing the disc face-on but unresolved. 
		
\begin{figure}
		\resizebox{\hsize}{!}{\includegraphics[scale=0.5,clip=true,trim=1.5cm 0.6cm 4.4cm 17cm]{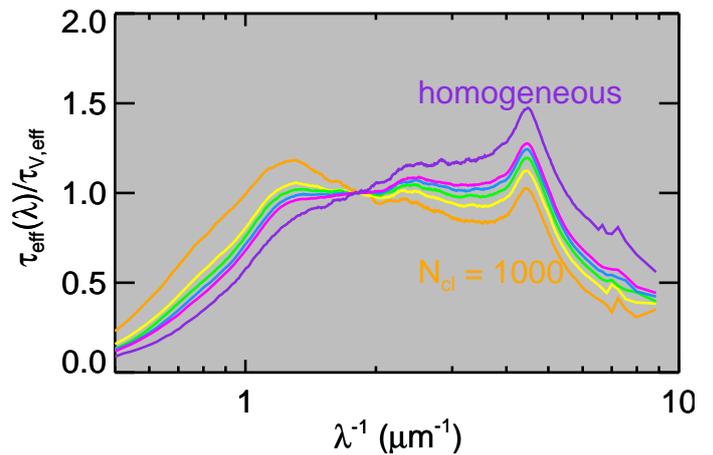}}
		\caption{Effective extinction curves of face-on discs with an opening angle of 45$\degr$, computed for aCSi dust. The colours are again the same as in Fig. \ref{fig:SaCcurves}, with orange indicating $N_{\rm cl} = 1000$, and proceeding through yellow, green and blue for $N_{\rm cl} = 2000, 3000, 5000$ to magenta specifying $N_{\rm cl} = 10000$, with violet indicating a homogeneous disc. The feature at $\sim 4-5\mathrm{\mu m^{-1}}$ does not change while the continuum scattering shows significant changes in the optical and UV.}
		\label{fig:scacurves}
\end{figure}

		As the observer's view of the star is unobstructed, the effective optical depth is negative for all wavelengths, due to the addition of the scattered photons to the stellar emission. 
		The wavelength dependence of this negative extinction(= scattered flux) can be interpreted to yield information concerning the scattering properties of the dust. 
		However, as seen in Fig. \ref{fig:scacurves}, the presence of clumpy structure alters the wavelength dependence of the scattered light, making the deduction of the scattering properties an unreliable process.
		Although Fig.~\ref{fig:scacurves} shows only one disc opening angle (in this case 45\degr), the same behaviour is seen for all opening angles between 5 and 45\degr.

		It is clear that as the clumps become increasingly optically thick, the scattered light in the UV continuum is suppressed compared to the optical. 
		The strong peak at $\sim$ 2000\,\AA\,that is visible in Fig. \ref{fig:scacurves} should not be confused with the 2175\,\AA\,extinction bump. 
		It is created by scattering, coincides with the maximum of $K_{\rm{sca}}$ (see Fig.~\ref{fig:ksca}) and is not affected by the UV absorption. 
		Therefore the feature is unaffected  by the optical depth of the clumps.
		Conversely, because of stronger absorption in the optical and UV, the scattered flux in the NIR domain is enhanced relative to the optical.
		Due to their clumpy structure, there may be unobstructed sight-lines to regions deep within the disc.
		Clumps at such locations can then scatter photons into the observer's line of sight, but before escaping may encounter further clumps.
		Since the clumps are optically thick at shorter wavelengths, they would preferentially absorb optical/UV photons, while the NIR photons have a significantly higher escape probability.
		Since the strength and wavelength of the aforementioned scattering peak are functions of the size and chemical composition of the dust grains \citep[see e.g.][]{2009ApJ...696.1502H} and highly model dependent, it may be possible to infer the degree of clumpiness of a disc with sufficiently precise measurements of the integrated scattered flux at NIR, optical and UV wavelengths, and comparing the shape of the scattered continuum to any features observed.
	
		If the relative enhancement of infrared scattering continues at wavelengths as long as $5\mu\mbox{m}$ then it may represent a source of contamination to the observations of ``core-shine'' \citep{2010A&A...511A...9S}, which are used to infer the presence of large grains in molecular cloud cores.
		\citet{2010A&A...511A...9S} report excess scattered flux in \textit{Spitzer}/IRAC \citep{2004ApJS..154...10F} bands 1 and 2 (at 3.6 and 4.5$\mu\mbox{m}$, respectively) toward dense regions ($A_\mathrm{V} \geq 10$), while the longer wavelength bands show only absorption in the most dense parts of the clouds.
		In principle, it is possible that dense clouds consist of many small, dense clumps that are not resolved in the observations, and if so the effective scattering behaviour would be modified, resulting in the apparent increase in infrared scattering.
		
		{ 
		Similarly to Sect. \ref{sec:shell}, we can also explore the effect of extinction when the star is viewed through the disc.
       Different optical paths through the disc will have radically different covering fractions of clumps, with paths through the mid-plane fully covered and lower covering fractions when the disc is viewed at lower inclination angles.
       
       As expected, the extinction curve through an edge-on clumpy disc exhibits the same behaviour as that for a clumpy shell \citep[e.g.][]{2015arXiv150804343S}. 
       This remains the same as long as the entire beam is within the disc (i.e. approximately when $i\geq 90-\alpha$).
       However, for grazing and near-grazing inclinations, the behaviour of the extinction curve becomes chaotic, due to the complexity of the scattered radiation field.
       This effect is a major concern for studies of extinction towards e.g. AGN tori, where the extinction seen through the torus will bear little resemblance to the wavelength dependence of the dust properties.
       Our results specifically indicate that studies which infer large grains in the circumnuclear medium \citep[e.g.][]{2001A&A...365...28M,2014ApJ...792L...9L} have to consider the possibility of significant contamination from radiative transfer effects.
		
		}
		
	\subsection{Extinction in a clumpy diffuse ISM\label{sec:ism}}

		The ISM is believed to be a highly turbulent, inhomogeneous medium with structure on all scales in both the dense and diffuse phases \citep[e.g.][]{1994ApJ...423..681V,1997ApJ...474..730P,1998PhRvE..58.4501P}.
		It is therefore interesting to consider whether the effect of clumps on extinction as described above in Sects. \ref{sec:shell} and \ref{sec:disc} also influence extinction in the diffuse galactic ISM.
		
{		The previous two subsections have examined scenarios in which star and dust are co-located relative to the observer, but to examine the influence of scattering on extinction in the diffuse galactic ISM it is interesting to consider scenarios where the dust is distributed along the entire line of sight between the observer and the star.
		
		To explore this, we model the ISM as a cuboid viewed along its long axis. 
		This cuboid is homogeneously filled with dust, such that the optical depth in V-band from the observer to the star ranges from 0.3 -- 20. 
		Since we reproduce typical ISM optical depths, the interaction probability for each radiation packet is small.
		As a result, we must run large numbers of packets ($\sim 10^{9}$) to achieve good statistics.
		To include the influence of back-scattered as well as forward-scattered photons, 5\% of the model volume is behind the star as seen from the observer. 
		The extinguished star is assumed to be at a distance of $100\mbox{ pc}$, however as this is a resolution dependent effect the model space can be uniformly rescaled to greater distances. 
		The model cuboid is scaled so that the cross-section is 50\arcsec.
		We then solve the radiative transfer and generate images of both the scattered and emitted radiation by ray-tracing as outlined in Sect.~\ref{sec:MC}. 
		From the images we extract a $5\arcsec\times5\arcsec$ aperture in order { to approximately match the diffraction limited resolution} of IUE, which remains the major source of UV data for extinction. }
		
		Contrary to the clumpy screen models of \citet{2005ApJ...619..340F,2011A&A...533A.117F} we include the effect of back-scattering by embedding the source within the dust column.
		Instead of a homogeneous density distribution, clumps are distributed randomly throughout the model space with fixed volume filling factor $f_{\mathrm{V}}\sim$1.5\% \citep[Eq. 27, assuming a two-phase ISM with the properties of the local ISM given therein]{2003ApJ...587..278W} 
		so that the free parameters are the total dust mass and clump number.
		The clump radii ${R_\mathrm{cl}}$ are calculated from the number and filling factor of clumps, such that all the clumps within each model are identical i.e. \begin{equation}
		{R_\mathrm{cl}} = \left[\frac{3 \ f_{\mathrm{V}} \times x \times y \times z}{4 \pi \ N_\mathrm{cl} }\right]^{1/3} 
		\end{equation} where $x,y,z$ are the dimensions of the model cuboid. 
		As the clumps are randomly distributed, we vary the random seed to explore the parameter space created by the variations in the positions of the clumps.

		\begin{table}
		\caption{Clumpy ISM model parameters}
		\label{tab:ISMpar}
		\centering
		\begin{normalsize}
		\begin{threeparttable}
		\begin{tabular*}{\hsize}{l l l l}
		\hline\hline
		\multicolumn{3}{c}{Parameter}   &   Values  \\ \hline
		&\\
		Dust mass & [$M_{\odot}$] & $M_\mathrm{d}$ & 0.0058, 0.007, 0.009, \\
		&& & 0.0115, 0.035, 0.058, \\
		&& &  0.07, 0.09, 0.115,  \\
		&& &  0.35, 0.58  \\
		Optical depth& (GraSi) & $\tau_\mathrm{V}$\tnote{a} & 0.16, 0.2, 0.25  \\
		&& & 0.3, 1.0, 1.6,  \\
		&& & 1.9, 2.5, 3.2,  \\
		&& &  9.7, 16  \\
		Clump number && $N_\mathrm{cl}$ & 10, 50, 100, 500  \\
	
		\hline
		\end{tabular*}
		\begin{tablenotes}

		\item [a] Optical depths (measured in the V band) of the homogeneous models of the respective dust masses.

		\end{tablenotes}
		\end{threeparttable}
				\end{normalsize}
		\end{table}
		
\begin{figure}[t]
			\resizebox{\hsize}{!}{\includegraphics[clip=true,trim=12cm 8cm 5cm 8cm]{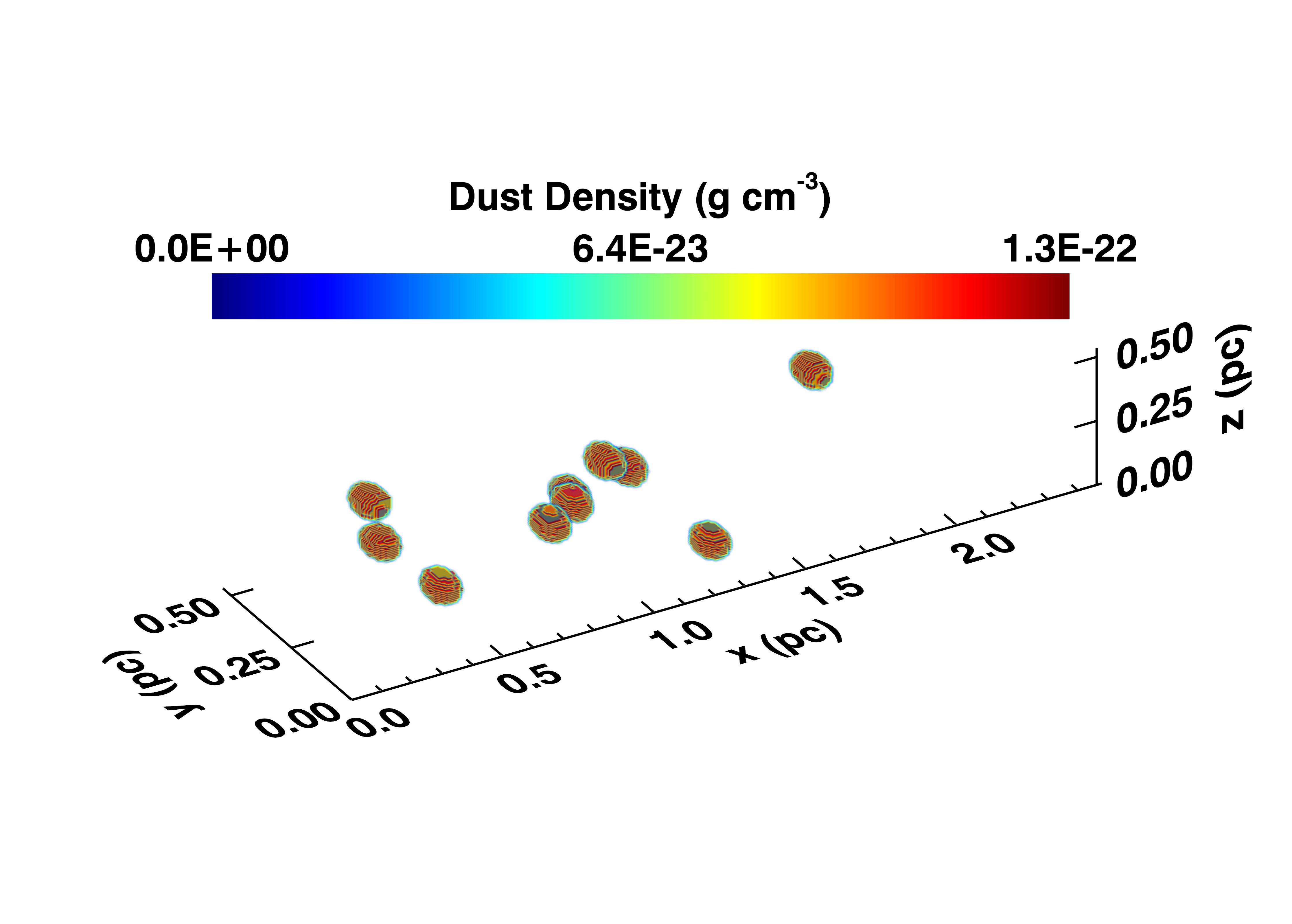}}

			\caption{Example section from a 1-phase clumpy density distribution (spherical clumps of constant density). The 2-phase distributions appear identical albeit with the intraclump space filled with a diffuse medium $10^{-4}$ times less dense than the clumps. This figure is constructed in a similar manner to Figs. \ref{fig:shellclumps}, \ref{fig:discclumps}, but the source is no longer within this section of the model due to the extreme length of the cuboid. In the ISM models, the source is not placed at the centre, but at a point 95\% of the length of the model cuboid. This section is that closest to the observer.}
			\label{fig:1pclumps}
		\end{figure}

\begin{figure}[!t]
			\resizebox{\hsize}{!}{\includegraphics[clip=true,trim=7cm 16cm 0cm 12cm]{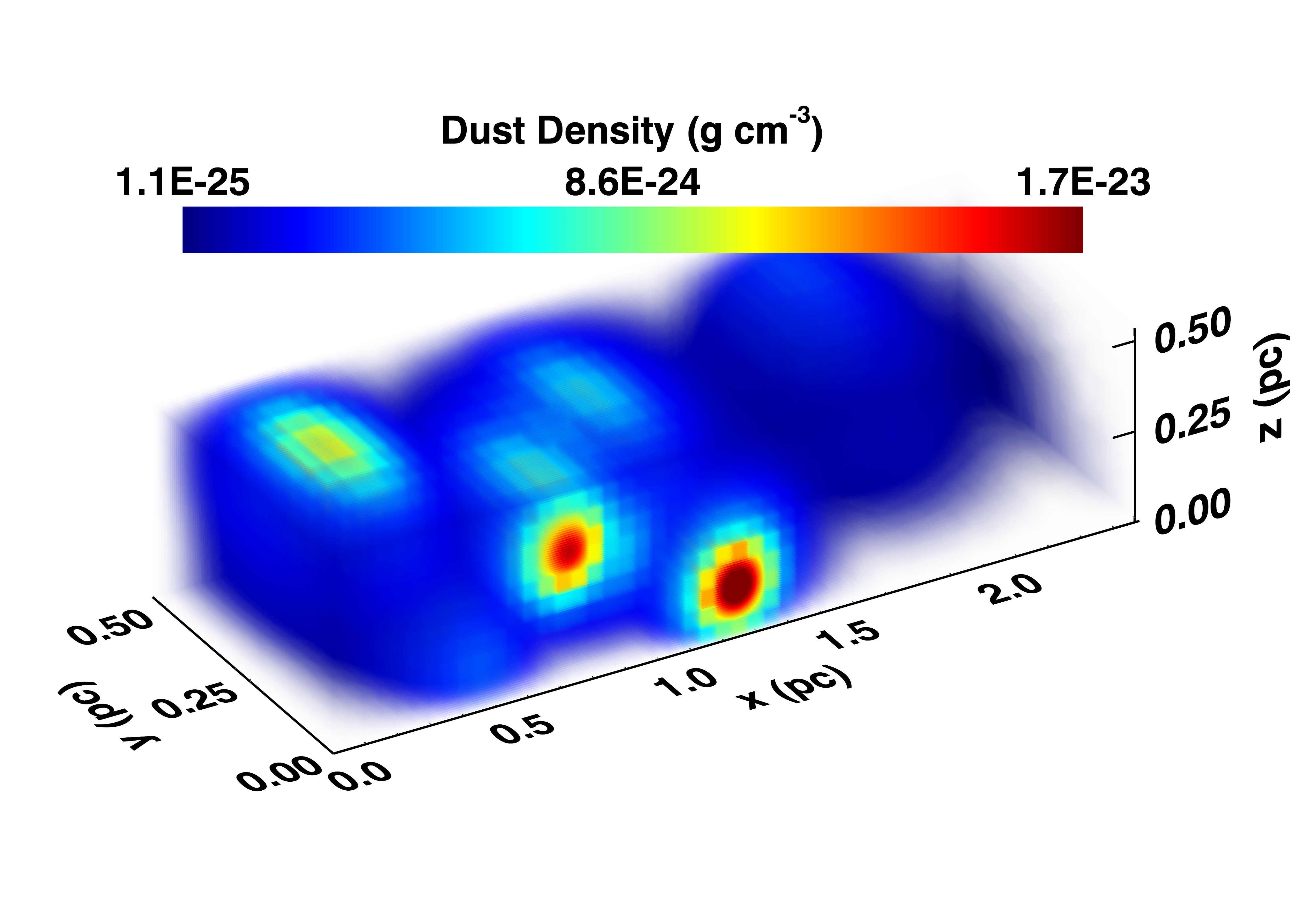}}
			
			\caption{As in Fig. \ref{fig:1pclumps} but showing an example section from a pressure-constrained clump density distribution.}
			\label{fig:pcclumps}
		\end{figure}

		We consider three different simple descriptions of the clumps in these geometries: 
		\begin{enumerate}
		\item Spherical clumps (1-phase) i.e. $\rho\left(R\right) = \rho_{\mathrm{cl}}$ for $R\leq R_{\mathrm{cl}}$, 0 elsewhere;
		\item as above but with the clumps embedded in a diffuse medium (2-phase) i.e. $\rho\left(R\right) = \rho_{\mathrm{cl}}$ for $R\leq R_{\mathrm{cl}}$, $10^{-4}\rho_{\mathrm{cl}}$ elsewhere;
		\item Pressure-constrained isothermal clumps with \begin{equation}\frac{\rho\left(R\right)}{\rho_{0}} = \frac{1}{1+\left(\frac{R}{R_\mathrm{cl}}\right)^2} \end{equation} to give a smoothly varying density distribution;
\end{enumerate}		 
		examples of which can be seen in Figs. \ref{fig:1pclumps}, \ref{fig:pcclumps}.
		

\begin{figure}
\resizebox{\hsize}{!}{\includegraphics[scale=0.5,clip=true,trim=1.3cm 0.5cm 4.5cm 5cm]{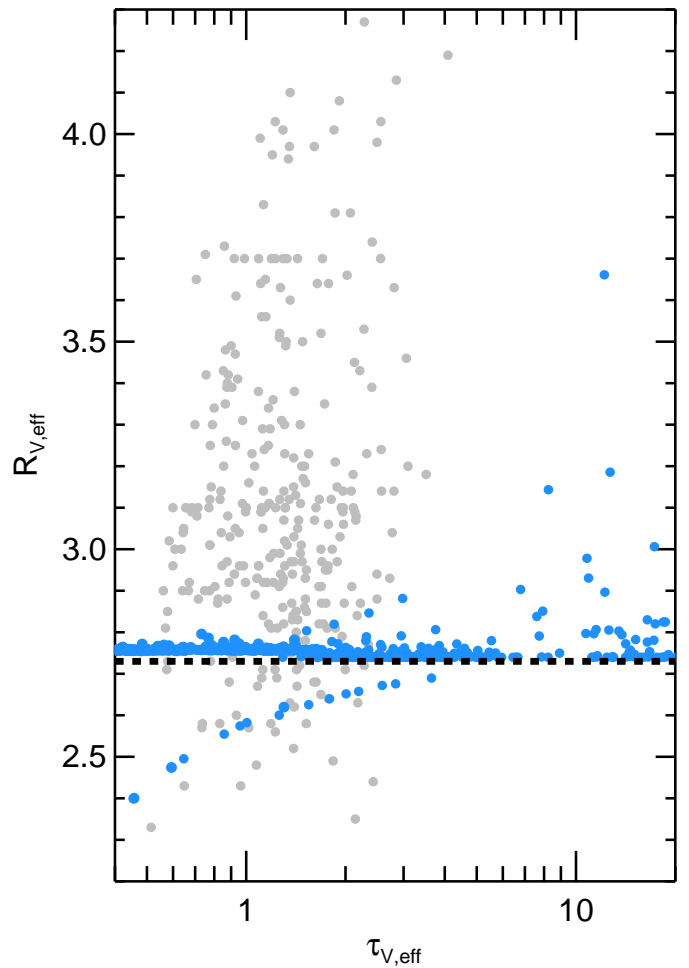}}

\caption{The range of $\Rve$ produced by clumpy ISM models. Our models are indicated by filled blue circles. These are compared with the observed values of $R_{\mathrm{V}}$ reported in \citetalias{2007ApJ...663..320F} (gray full circles) taking $\tau_{\rm V,eff} = A_{\rm V} / 1.086$. The thick black dashed line indicates the $R_\mathrm{V}$ of the dust cross-sections. Except for a small fraction of outliers, which increases toward large optical depth, clumpiness has little effect on typical interstellar extinction curves.
}
\label{fig:RVrange}
\end{figure}		
		
{		The results from the clumpy ISM models can be seen in Fig. \ref{fig:RVrange}. 
       With the exception of a small fraction of outliers\footnote{Approximately consistent with the expected number of cases where a clump is close to the star.}, the effect of clumpiness on extinction is negligible except at high optical depth.
       This suggests that on lines of sight that avoid the galactic centre, the effect of scattering can be neglected, and extinction can be reliably be used as a probe of the properties of interstellar dust, as expected from \citet{1983ApJ...270..169P}.
       The fact that the OB stars typically used to measure extinction tend to clear a large volume (several pc) surrounding them of interstellar matter through wind and radiation pressure further reduces the probability that interstellar extinction is significantly modified by scattering on distance scales of a few kpc.
}

\section{Discussion}

		From Sect. \ref{sec:shell} it is clear that if the dust is concentrated in optically thick clumps, the extinction curve is artificially flattened.
		This has been previously highlighted by \citet{1984ApJ...287..228N}, however, they did not attempt to derive a relationship between the flattening of extinction and the clump properties.
		
\begin{figure}
			\resizebox{\hsize}{!}{\includegraphics[scale=0.5,clip=true,trim=1.5cm .5cm 4.5cm 16.5cm]{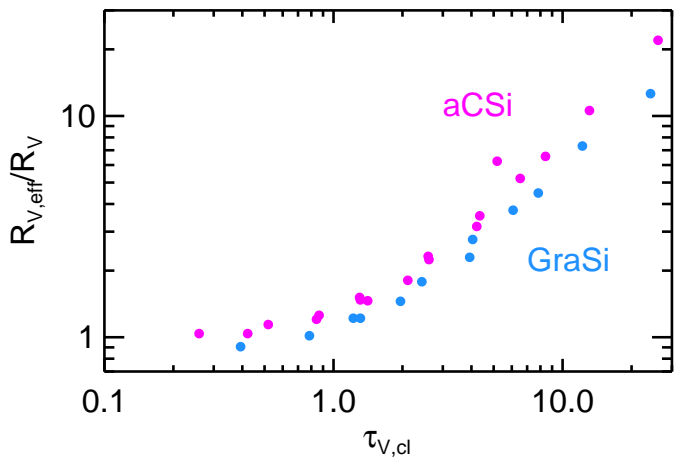}}

			\caption{Evolution of the effective extinction curve in clumpy shells with the optical depth of the clumps for both dust models. While the two models show generally the same behaviour, there remains an offset ($\sim$ 20\%) between the two.}
			\label{fig:taurv}
		\end{figure}

		Figure \ref{fig:taurv} demonstrates the relation between this increase in the V-band optical depth of the clumps and $\Rve$, using the results from Sect. \ref{sec:shell} for both dust models {  using all curves shown in figures~\ref{fig:SaCcurves}~\&~\ref{fig:GraScurves}}.
		While it is clear that both dust models follow similar trends, they appear to form two separate sequences.
		{ The separation between these sequences is typically a few tens of percent.} 
		
\begin{figure}
			\resizebox{\hsize}{!}{\includegraphics[scale=0.5,clip=true,trim=1.5cm .5cm 4.5cm 16.5cm]{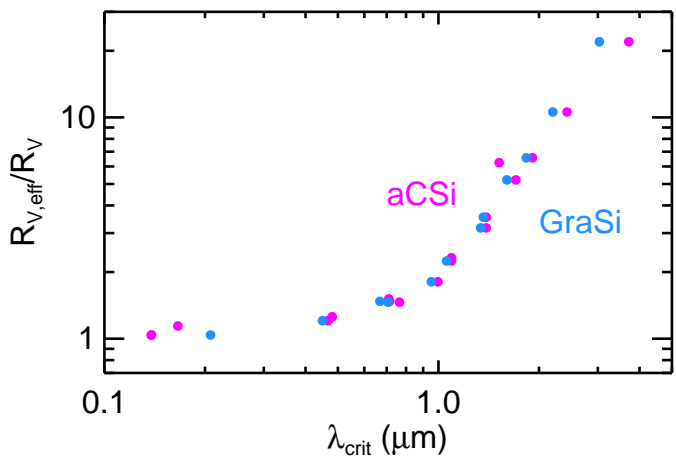}}

			\caption{As in Fig. \ref{fig:taurv}, but now comparing $R_\mathrm{V}$ to $\lambda_\mathrm{crit}$, the wavelength at which $\tau_\mathrm{cl} = 1$. The two previously disparate curves are now reconciled.}
			\label{fig:lamrv}
		\end{figure}

		Since the optical depth at an arbitrary wavelength is not directly related to the extinction curve, we wish to transform this to a quantity that is.
		As the changes in the shape of the extinction curve result from wavelength-dependent optical depth effects, we introduce the quantity $\lambda_\mathrm{crit}$, which is defined as the wavelength for which $\tau_\mathrm{cl}\left(\lambda_\mathrm{crit}\right)=1$.
		When the change in $R_\mathrm{V}$ is plotted against this critical wavelength (Fig. \ref{fig:lamrv}) the two dust models overlap.
		The gradient is rather shallow for $\lambda_\mathrm{crit}\leq 600\,\mbox{nm}$ but steepens dramatically beyond this.
		As $\Rve$ is related to the B- and V-band optical depths, it stands to reason that clumps that are optically thin or only marginally optically thick to these wavelengths would only weakly affect the shape of the extinction curve, while clumps that are optically thick at even longer wavelengths will have a much stronger effect.
		
		If the true $R_\mathrm{V}$ of the dust can be determined independently of the extinction measurements, then it is in principle possible to use the relation between $\Rve$ and $\lambda_\mathrm{crit}$ to infer the structure of the medium.
		
		{ It is important to note that all the effects described in this paper will become more significant as the distance between the object and the observer increases, as structure will be ever more poorly resolved.
		Therefore, extragalactic observations are particularly susceptible, and even more so at high redshift.
		This means that studies that use AGN \citep[e.g.][]{2001A&A...365...28M} or, in particular GRBs \citep{2011A&A...532A.143Z} to probe dust properties must be especially careful to consider the role of radiative transfer effects.
		}
		
		\section{Summary}

		Clumpy media and the collection of scattered light can fundamentally alter the observed extinction curve, which can significantly hinder the accurate interpretation of observations{  , in particular when the scattering medium is close to the extinguished object}.
		In clumpy media the changes {in the shape of the extinction curve} are related not to the optical depth of the clumps but rather to the critical wavelength for which $\tau_\mathrm{cl}\left(\lambda_\mathrm{crit}\right)=1$.
		If the true $R_\mathrm{V}$ is known, it is in principle possible to infer the structure of the medium from this relationship.
{		Furthermore, there is a shift in the wavelength of the peak of the 2175\,\AA\, feature towards shorter wavelengths.}
		
		Similarly, the observed scattering behaviour of dust can be markedly different if the scattering medium is clumpy rather than homogeneous.
		More optically thick clumps lead to a suppression of the optical and UV scattered flux in the continuum, while scattering features are unaffected, potentially providing a means by which to constrain the structure of a scattering medium.
		
		We have shown that the collection of scattered photons represents a major challenge to measurements of extinction towards embedded objects, { particularly in other galaxies e.g. AGN or GRBs, where large-scale structure is unresolved}. 
		As a result, there is not necessarily a 1:1 link between the extinction curve and the wavelength dependence of dust cross-sections.
		{  However, the effect on observations of diffuse galactic extinction is negligible.}

\begin{acknowledgements}
{
We thank the anonymous referee and the editor whose comments helped improve this manuscript.
We thank Endrik Kr\"{u}gel for helpful discussions and for providing the original version of his MC code, and Frank Heymann for discussions on the implementation of perspective-projection ray tracing.
We are grateful to Sebastian Wolf for discussions, comments and suggestions which improved the content of this manuscript.
PS is supported under DFG programme no. WO 857/10-1
}
\end{acknowledgements}

\bibliographystyle{aa} 
\bibliography{clumpy}

\end{document}